# Anomalous optical coupling between two silicon wires of a slot waveguide in epsilon-near-zero metamaterials

Jie Gao[†] and Xiaodong Yang[‡]

*Department of Mechanical and Aerospace Engineering,*

*Missouri University of Science and Technology, Rolla, Missouri 65409, USA*

[†]gaojie@mst.edu, [‡]yangxia@mst.edu

**Abstract:** Anomalous optical coupling properties between two silicon wires in a silicon slot waveguide embedded in epsilon-near-zero (ENZ) metamaterials are proposed and demonstrated. The dependences of optical field enhancement in the slot region and transverse optical force on the slot size and the permittivity of surrounding material are studied in details. It is demonstrated that the optical field in the slot region is significantly enhanced due to the giant index contrast at the slot interface between silicon wires and ENZ metamaterials, but the optical mode coupling between silicon wires is greatly reduced so that the transverse optical force is suppressed into almost zero. Moreover, metal-dielectric multilayer structures are designed to realize ENZ metamaterials in the slot region for achieving the electric field enhancement.

## 1. Introduction

The emergence of metamaterials with artificially engineered subwavelength structures offers a flexible method to control the permittivity and permeability of materials. With metamaterials, many exotic optical phenomena have been demonstrated, such as negative index of refraction [1], ultrahigh index of refraction [2], and epsilon-near-zero (ENZ) materials [3]. Due to the anomalous electromagnetic properties, metamaterials with near-zero permittivity have been widely investigated in both theory and engineering [4-7]. Maxwell's equations state that, for a high-index-contrast interface, the continuity of the normal component of electric displacement will result in a considerable electric field enhancement at the low-index region. Conventional silicon slot waveguides embedded in low-index materials have been widely studied to enhance the optical field and confine light in the slot region [8]. In order to further boost the electric field enhancement, either the index of the waveguide material can be increased, or the index of the slot material can be reduced. Due to the ultrahigh refractive index of hyperbolic metamaterials, metamaterial slot waveguides embedded in air have been proposed to achieve strong optical field enhancement [9]. Another concept is to use ENZ metamaterials as the low-index material in order to achieve the ultra-large index contrast, which has been used to realize anomalous field enhancement [10, 11].

In this paper, we present a new type of silicon slot waveguides embedded in ENZ metamaterial surroundings, where the electric field in the slot region can be significantly enhanced, due to the giant index contrast at the slot interface. The dependences of the electric field enhancement in the slot region with the slot size and the permittivity of surrounding material are systematically studied. Furthermore, it is also demonstrated that the transverse optical force between two silicon wires of silicon slot waveguides with different slot sizes will keep almost zero, due to the weak mode coupling of silicon wires in the ENZ metamaterial surroundings. Such extraordinary properties of silicon slot waveguides embedded in ENZ metamaterial surroundings will be very useful for enhanced light-matter interactions, such as nonlinear optics [12], sensitive mechanical sensors [13], and optomechanical device actuation [14].

## 2. Electric field enhancement

Figure 1(a) shows the schematic of the silicon slot waveguides, where two silicon wires with square cross sections are closely placed in the isotropic ENZ environment with a nanoscale slot size g along $x$ direction. The refractive index of silicon is $\varepsilon_{si} = 3.482$. By assuming that $a = 350$ nm and $g = 2$ nm, Figs. 1(c) and 1(d) show the calculated electric field distributions at $\lambda_0 = 1.55$ $\mu$m with the ENZ metamaterial ($\varepsilon_{surr} = 10^{-4}$) and air ($\varepsilon_{surr} = 1$) surroundings, respectively. Considerable electric field enhancement can be observed at the slot region in both cases. Since the gap is very small, mode coupling exists between the two silicon wires. The leaked electric field from the silicon wires will be greatly enhanced in the slot region due to its lower permittivity. However, it can be clearly seen that ENZ metamaterial surroundings can lead to a much stronger electric field enhancement. This is due to the giant permittivity contrast between the silicon wire and the slot region, so that even weak optical field leakage from the silicon wires can result in strong electric field enhancement. The symmetric mode profiles of the silicon slot waveguides with slot size $g = 10$ nm in the ENZ metamaterial surroundings ($\varepsilon_{surr} = 10^{-4}$) are shown in Fig. 2. It is shown that strong electric field $E_x$ is localized inside the slot region, since the leaked electric field from the silicon wires can be greatly enhanced by the low permittivity in the gap. While the magnetic field $H_y$ is tightly confined in the silicon wires, as shown in Fig. 2(b). Accordingly, there is strong optical energy flow guided in the slot region, as shown in Fig. 2(c).

The electric field enhancement factor $\eta$ is defined as the ratio of the electric field $E_x$ at the two boundaries of one silicon wire $\eta = E_x[|x| = (g/2)^-]/E_x[|x| = (g/2+h)^+]$, which represents the electric field enhancement in the slot region. The simulated effective refractive index along the propagation direction $n_{eff,z}$ and the electric field enhancement factor $\eta$ as a function of slot size $g$ for two different surrounding materials are shown in Figs. 3(a) and 3(b). It is shown in Fig. 3(a) that the effective refractive index of the slot waveguides with air surroundings gradually increases as gap size gets smaller. On the other hand, the effective refractive index of the slot waveguides with ENZ metamaterial surroundings is almost unchanged, indicating that the two silicon wires are optically insulated to each other by the ENZ material and the mode coupling between the silicon wires is quite weak. However, as shown in Fig. 3(b), the electric field in the ENZ slot can be greatly enhanced at the small gap size, which is much stronger than the case of air slot. Figures 3(c) and 3(d) present the simulated effective refractive index $n_{\text{eff},z}$ and the electric field enhancement factor $\eta$ as a function of the permittivity of surrounding material $\varepsilon_{\text{surr}}$ for $g = 2$ nm and $g = 5$ nm. As the permittivity of surrounding material gets smaller, the effective refractive index will get lower and reach almost a constant when $\varepsilon_{\text{surr}}$ is less than $10^{-2}$. At the same time, as the permittivity of surrounding material gets smaller, the electric field enhancement will get stronger and then keep unchanged. It is noted that as $\varepsilon_{\text{surr}}$ gets smaller, the optical mode coupling between these two silicon wires becomes weaker.

In order to provide a comprehensive understanding of the mechanism of optical coupling between two silicon wires in slot waveguides with ENZ metamaterials, theoretical analysis is conducted. Since the optical mode profiles in Fig. 2 show a negligible dependence on the $y$ direction, the 3D silicon slot waveguides can be approximately treated as 2D slot waveguides, as shown in Fig. 1(b). Assuming the symmetric optical mode profiles of the slot waveguides have the form of $\exp(i\beta z - i\omega t)$, the electric field $E_x$ can be expressed as

$$E_x = E_0 \begin{cases} \dfrac{1}{\varepsilon_{\text{surr}}} \cos\left(-k_x \dfrac{a}{2} + \varphi\right) \dfrac{\cosh(\gamma x)}{\cosh(\gamma g/2)}, & 0 \leq |x| \leq \dfrac{g}{2} \\ \dfrac{1}{\varepsilon_{\text{si}}} \cos\left[k_x\left(|x| - \dfrac{a+g}{2}\right) + \varphi\right], & \dfrac{g}{2} \leq |x| \leq \dfrac{g}{2} + a \\ \dfrac{1}{\varepsilon_{\text{surr}}} \cos\left(k_x \dfrac{a}{2} + \varphi\right) \exp\left[-\gamma\left(|x| - \dfrac{2a+g}{2}\right)\right], & |x| \geq \dfrac{g}{2} + a \end{cases} \qquad (1)$$

Here, $\beta$ is the mode propagation constant, $\omega$ is the angular frequency at $\lambda_0 = 1.55$ $\mu$m, $\varphi$ is the phase shift at the middle of each silicon wire due to the mode coupling. The wave vector in the silicon wires $k_x$ and the field decay rate $\gamma$ in the surrounding material are related to $\beta$ through the dispersion relations $k_x^2 + \beta^2 = \varepsilon_{\text{si}} k_0^2$ and $\beta^2 - \gamma^2 = \varepsilon_{\text{surr}} k_0^2$ for silicon wires and surrounding materials, respectively. According to the continuity conditions of the tangential field components at the silicon wire interface, the following equations can be obtained

$$\begin{cases} \tan\left(-k_x \dfrac{a}{2} + \varphi\right) = -\dfrac{\gamma \varepsilon_{\text{si}}}{k_x \varepsilon_{\text{surr}}} \tanh\left(\gamma \dfrac{a}{2}\right) \\ \tan\left(k_x \dfrac{a}{2} + \varphi\right) = \dfrac{\gamma \varepsilon_{\text{si}}}{k_x \varepsilon_{\text{surr}}} \end{cases}. \qquad (2)$$

Therefore, the mode propagation constant $\beta$ and corresponding optical field properties can be obtained by solving the above equations. The theoretical analysis results of the effective refractive index and electric field enhancement factor are plotted in Fig. 3. As shown in Fig. 3(a), although the effective refractive index derived from the 2D theoretical analysis is slightly different from the 3D numerical simulation, the trends are almost the same. The analytical expression for the enhancement factor $\eta$ can be derived with Eqs. (1) and (2),

$$\eta = \frac{\cos(-k_x a/2 + \varphi)}{\cos(k_x a/2 + \varphi)} \approx \sqrt{\frac{(n_{si}^2 - n_{eff}^2)n_{surr}^4 + (n_{eff}^2 - n_{surr}^2)n_{si}^4}{(n_{si}^2 - n_{eff}^2)n_{surr}^4 + (n_{eff}^2 - n_{surr}^2)n_{si}^4 k_0^2 (g^2/4)}} \quad (3)$$

When $n_{surr} \to 0$ or $n_{eff} \gg n_{surr}$, Eq. (3) can be simplified into $\eta \approx 2/(gk_0 n_{eff})$. For the silicon slot waveguides with ENZ metamaterial surroundings, $n_{eff}$ does not change as $g$ get smaller so that $\eta$ will increase quickly when $g$ gets small (< 5 nm). When the silicon slot waveguides are surrounded by the ENZ metamaterial, the strong discontinuity of permittivity at the silicon wire interface will result in the weak mode coupling between the silicon wires due to the fact that the optical energy is very hard to leak into the ENZ metamaterial surroundings. However, the leaked electric field can still be greatly enhanced due to the near-zero permittivity of the ENZ metamaterial, as shown in Fig. 3(b). On the other hand, for the case of the silicon wires in air, the optical energy leaked from the silicon wire interface is strong so that the enhanced electric field can be achieved.

## 3. Optical forces between the silicon wires

The optical coupling between the silicon wires with ENZ metamaterial surroundings can be analyzed with the coupled mode theory [15]. The eigenmode supported by a slot waveguide system can be treated as the superposition of two individual silicon wire modes and the coupling strength between the two identical silicon wires can be derived as $\kappa = n_{eff,z} - n_{0,z}$, here $n_{eff,z}$ is the effective refractive index of the coupled silicon wires and $n_{0,z}$ is the effective refractive index of the single silicon wire. For the silicon slot waveguides with ENZ metamaterial surroundings, the coupling strength $\kappa_{ENZ}$ can be derived through Eq. (2) at the condition that $\varepsilon_{ENZ} \ll 1$,

$$\kappa_{ENZ} = \frac{\pi \lambda}{2a^3 \varepsilon_{si} \gamma_0^2} \left[ \frac{1}{\tanh(\gamma_0 g/2)} - 1 \right] \varepsilon_{ENZ} \quad (4)$$

Here, $\gamma_0$ is the field decay rate of single silicon wire in the ENZ metamaterial surroundings ($\varepsilon_{surr} = 0$), $\gamma_0 = \sqrt{\varepsilon_{si} k_0^2 - (\pi/a)^2}$. It shows that the coupling strength between the silicon wires is proportional to the permittivity of surrounding material. For the silicon slot waveguides with ENZ metamaterial surroundings, $\kappa_{ENZ}$ is close to zero so that $n_{eff,z}$ is almost unchanged as the gap size gets smaller. At the same time, the energy will be tightly confined within the silicon wire, due to the large discontinuity of permittivity at the silicon wire interface. This is quite different from the silicon wires in air. The optical coupling strength in the slot waveguides is directly related to the transverse optical forces exerted on each silicon wire. In general, when the electrical field in the slot region is large, the transverse optical force between the two silicon wires is also very strong. However, in the silicon slot waveguides with ENZ metamaterial surroundings, it is found that the optical force between the two silicon wires is very small due to the weak mode coupling even though the optical field is greatly enhanced. The optical force generated on the silicon wires can be calculated by integrating the Maxwell's stress tensor $\bar{\bar{T}} = \varepsilon \vec{E}\vec{E} + \mu \vec{H}\vec{H} - \bar{\bar{I}}/2\left(\varepsilon_0|\vec{E}|^2 + \mu_0|\vec{H}|^2\right)$ around any arbitrary surface enclosing the silicon wire [15]. After substituting the optical fields into the Maxwell's stress tensor and taking into account that $n_{surr} \ll 1$, the following analytical expression for the optical forces on the slot waveguides can be obtained

$$f \approx \frac{2k_x}{c(1+ak_x)} \frac{\left(n_{si}^2 - n_{eff}^2\right)}{n_{si}^2 n_{eff}^3} \left(\frac{n_{surr}}{k_0 g}\right)^2 \tag{5}$$

From Eq. (5), it is noted that for a fixed $g$, the ENZ metamaterial surroundings will greatly reduce the optical force in the slot waveguides. As $n_{surr}$ is close to zero, the optical force will disappear. Figure 4 plots the optical force as functions of slot size $g$ and the permittivity of surrounding material $\varepsilon_{surr}$ for both of the numerical simulation and the theoretical analysis.

Figure 4(a) shows that the optical force increases significantly as the gap size shrinks when the surrounding material is air, which have been studied previously [16, 17]. On the other hand, the optical force will almost keep as zero when the slot waveguides are in the ENZ metamaterial surroundings, as predicted from Eq. (5). It shows that unlike the conventional slot waveguides in air, there is almost no optical force in the slot waveguides with ENZ metamaterial surroundings, although the optical field can be greatly enhanced in the ENZ slot region. This behavior can also be explained by the coupled mode theory. It has been indicated that for a slot waveguide system with two identical silicon wires, the transverse optical force can be written as [18]

$$f = \frac{1}{c}\left.\frac{\partial n_{eff,z}}{\partial g}\right|_\omega = \frac{1}{c}\left.\frac{\partial \kappa}{\partial g}\right|_\omega \tag{6}$$

It means that the transverse optical force is proportional to the variation rate of the effective refractive index or the coupling strength as the two silicon wires approach to each other adiabatically. For the case of the silicon wires with ENZ metamaterial surroundings, the mode coupling between the two silicon wires is quite weak. The effective refractive index almost does not change when $g$ gets smaller, so that the transverse optical force will keep almost zero. On the other hand, the effective refractive index of the silicon wires in air will gradually increase as the gap size shrinks, resulting in strong transverse optical forces. Figure 4(b) gives the optical force as a function of the permittivity of surrounding material $\varepsilon_{surr}$ for two different slot sizes. It can be seen that larger optical forces will be obtained for the smaller gap size for the same surrounding material. As the permittivity of surrounding material reduces from one, the optical field in the slot region will increase in the beginning, as shown in Fig. 3(d). This will cause the optical force to increase. As the permittivity of surrounding material gets much less than 0.1, the optical force will tend to be zero due to the greatly reduced mode coupling in the silicon slot waveguides, as indicated in Eq. (4).

## 4. Realistic multilayer structures

Silicon slot waveguides filled with realistic ENZ metamaterials can be constructed, as shown in the inset of Fig. 5(a). The alternative silver and germanium layers with the period of 15 nm are used to make the ENZ metamaterials. According to the effective medium theory (EMT), the multilayer metamaterials can be regarded as a homogeneous effective medium and the principle components of the permittivity can be expressed as follows [9]

$$\varepsilon_x = \varepsilon_z = f_m \varepsilon_m + (1 - f_m) \varepsilon_d$$
$$\varepsilon_y = \frac{\varepsilon_m \varepsilon_d}{f_m \varepsilon_d + (1 - f_m) \varepsilon_m} \quad (7)$$

where $f_m$ is the volume filling ratio of silver, $\varepsilon_d$ and $\varepsilon_m$ are the permittivity of germanium and silver, respectively. $\varepsilon_d = 16$, and $\varepsilon_m(\omega) = \varepsilon_\infty - \omega_p^2/(\omega^2 + i\omega\gamma)$ is based on the Drude model, with a background dielectric constant $\varepsilon_\infty = 5$, plasma frequency $\omega_p = 1.38 \times 10^{16}$ rad/s and collision frequency $\gamma = 5.07 \times 10^{13}$ rad/s. From Eq. (7), the permittivity tensor of hyperbolic metamaterials can be calculated as $\varepsilon_x = \varepsilon_z = 0.0078 + 0.6142i$, $\varepsilon_y = 18.3784 + 0.0136i$ for $f_m = 0.1146$ at $\lambda_0 = 1.55$ $\mu$m. Figure 5 plots the electric field $E_x$ profiles along $y = 0$ and $x = 0$ for $g = 20$ nm. In the exterior air surroundings and the silicon wires, the $E_x$ profiles calculated from the multilayer structures overlap with the EMT prediction. In the multilayer ENZ metamaterials slot region, the simulated $E_x$ profile will oscillate around along $y$ direction, as shown in Fig. 5(b), due to the coupling of gap plasmons between the multilayers. However, the electric field enhancement can be achieved by using the ENZ metamaterials with realistic multilayer structures.

## 5. Conclusions

In conclusion, we have proposed a new kind of silicon slot waveguides with ENZ metamaterial surroundings. The dependences of the electric field enhancement in the slot region and the transverse optical force between two silicon wires of a slot waveguide on the gap sizes and the permittivities of surrounding materials are investigated in both numerical simulation and theoretical analysis. It is demonstrated that although the optical field within the ENZ metamaterial slot region can be significantly enhanced due to the giant index contrast at the silicon wire interface, the mode coupling between the silicon wires is very weak and therefore the transverse optical force is suppressed into almost zero. The demonstrated results will open a new realm in many research areas in nanoscale light-matter interactions, such as nonlinear optics and optomechanics.

**Acknowledgments**

This work was partially supported by the Intelligent Systems Center and the Materials Research Center at Missouri S&T, the University of Missouri Interdisciplinary Intercampus Research Program, the University of Missouri Research Board, and the Ralph E. Powe Junior Faculty Enhancement Award. The authors acknowledge X. Jiao, Y. He and L. Sun for their discussions and help about this work.

**Figure captions**

**Figure 1.** Schematic of the silicon slot waveguides and the electric field enhancement in the slot region. (a) Three-dimensional (3D) silicon slot waveguide structure, and (b) the approximated two-dimensional (2D) slot waveguide structure for theoretical analysis. 3D surface plots of the $E_x$ field distributions [normalized to $E_x$ ($y = 0$, $|x|=|a + g/2|^+$)] for slot waveguides with (c) ENZ metamaterial surroundings and (d) air surroundings, respectively.

**Figure 2.** The optical mode profiles of (a) $E_x$, (b) $H_y$ and (c) $S_z$ for the silicon slot waveguides with $g = 10$ nm with ENZ metamaterial surroundings ($\varepsilon_{surr} = 10^{-4}$). The crossing line plots at $y = 0$ are also shown.

**Figure 3.** (a) The effective refractive index $n_{eff,z}$, and (b) the electric field enhancement factor $\eta$, as a function of slot size $g$ for ENZ metamaterial surroundings and air surroundings. (c) The effective refractive index $n_{eff,z}$, and (d) the electric field enhancement factor $\eta$, as a function of the permittivity of surrounding material $\varepsilon_{surr}$ for $g = 2$ nm and $g = 5$ nm. Results from both 3D finite-element method (FEM) numerical simulation (solid line) and 2D theoretical analysis (dashed line) are plotted.

**Figure 4.** The calculated transverse optical forces between two silicon wires in silicon slot waveguides with (a) different gap sizes $g$ and (b) different permittivities of surrounding materials $\varepsilon_{surr}$. Results from both 3D FEM simulation (solid line) and 2D theoretical analysis (dashed line) are plotted.

**Figure 5.** The electric field $E_x$ profiles [normalized to $E_x$ ($y = 0$, $|x|=|a + g/2|^+$)] at (a) $y = 0$ and (b) $x = 0$ for $g = 20$ nm. Results from both effective medium theory (EMT) and realistic metal-dielectric multilayer structures are plotted.

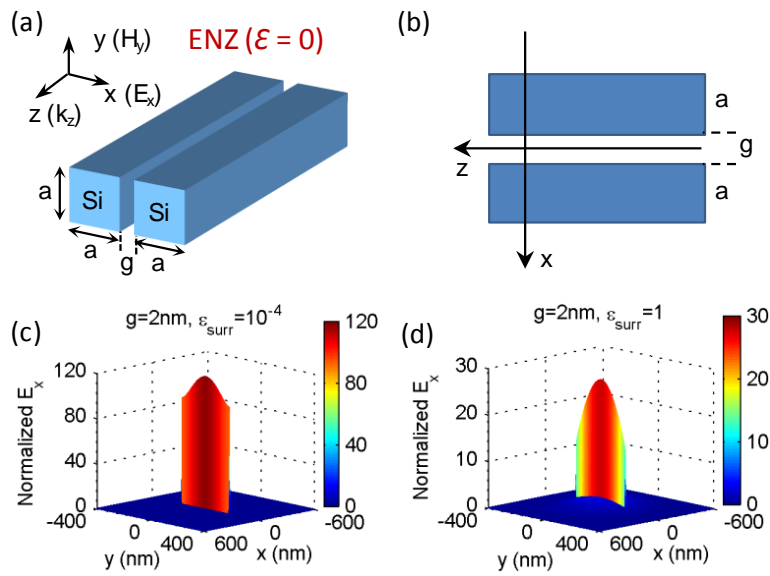

**Fig. 1.** Schematic of the silicon slot waveguides and the electric field enhancement in the slot region. (a) Three-dimensional (3D) silicon slot waveguide structure, and (b) the approximated two-dimensional (2D) slot waveguide structure for theoretical analysis. 3D surface plots of the $E_x$ field distributions [normalized to $E_x(y=0, |x|=|a+g/2|^+)$] for slot waveguides with (c) ENZ metamaterial surroundings and (d) air surroundings, respectively.

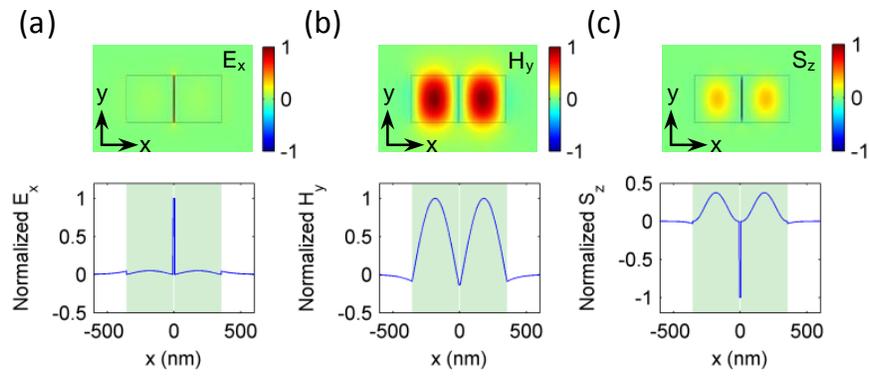

**Fig. 2.** The optical mode profiles of (a) $E_x$, (b) $H_y$ and (c) $S_z$ for the silicon slot waveguides with $g = 10$ nm with ENZ metamaterial surroundings ($\varepsilon_{surr} = 10^{-4}$). The crossing line plots at $y = 0$ are also shown.

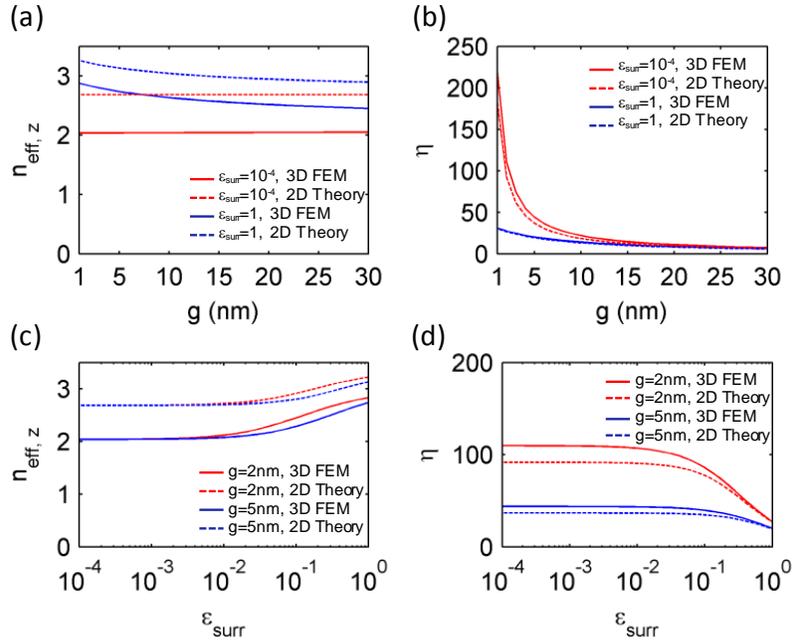

**Fig. 3.** (a) The effective refractive index $n_{eff,z}$, and (b) the electric field enhancement factor $\eta$, as a function of slot size $g$ for ENZ metamaterial surroundings and air surroundings. (c) The effective refractive index $n_{eff,z}$, and (d) the electric field enhancement factor $\eta$, as a function of the permittivity of surrounding material $\varepsilon_{surr}$ for $g = 2$ nm and $g = 5$ nm. Results from both 3D finite-element method (FEM) numerical simulation (solid line) and 2D theoretical analysis (dashed line) are plotted.

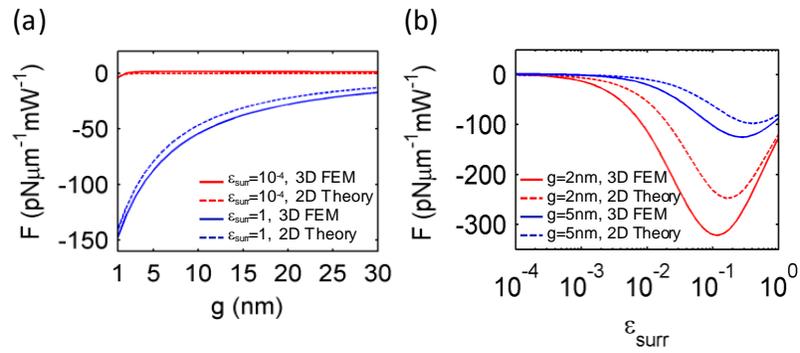

**Fig. 4.** The calculated transverse optical forces between two silicon wires in silicon slot waveguides with (a) different gap sizes $g$ and (b) different permittivities of surrounding materials $\varepsilon_{surr}$. Results from both 3D FEM simulation (solid line) and 2D theoretical analysis (dashed line) are plotted.

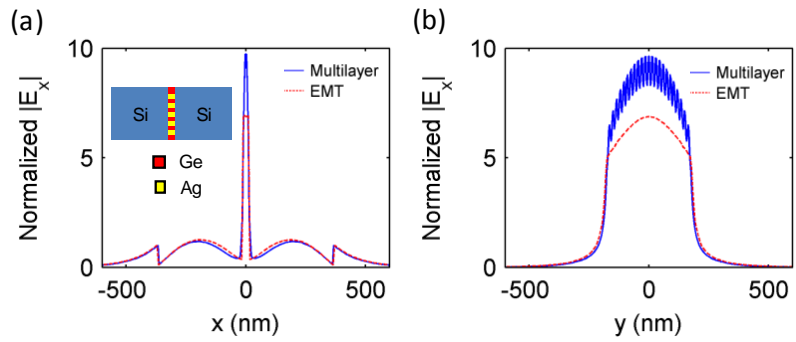

**Fig. 5.** The electric field $E_x$ profiles [normalized to $E_x$ ($y = 0$, $|x|=|a + g/2|^{+}$)] at (a) $y = 0$ and (b) $x = 0$ for $g = 20$ nm. Results from both effective medium theory (EMT) and realistic metal-dielectric multilayer structures are plotted.